\documentclass[reprint, aps,]{revtex4-1}
\usepackage{amssymb}
\usepackage{amsmath}
\usepackage{graphicx}
\usepackage{dcolumn}
\usepackage{bm}

\begin{document}

\title{Quantum interferometry via a coherent state mixed with a photon-added
squeezed vacuum state}
\author{Shuai Wang$^{1,2 \dagger }$, Xuexiang Xu$^{3}$ , Yejun Xu$^{4}$,
Lijian Zhang $^{2}$}
\address{$^1$ School of Mathematics and Physics, Jiangsu University of Technology, Changzhou 213001, P.R. China
\\$^{\dagger }$ Corresponding author: wshslxy@cczu.edu.cn}
\address{$^2$
College of Engineering and Applied Sciences, Nanjing University, Nanjing
210093, P.R. China}
\address{$^3$  Department of Physics, Jiangxi Normal University,
Nanchang 330022,P.R. China}
\address{$^4$ School of Mechanical and
Electronic Engineering, Chizhou University, Chizhou 247000, P.R. China}
\begin{abstract}
We theoretically investigate the phase sensitivity with parity detection on
a Mach-Zehnder interferometer with a coherent state combined with a
photon-added squeezed vacuum state. When the phase shift approaches zero,
the squeezed vacuum state is indeed the optimal state within a constraint on
the average number of photons. However, when the phase shift to be estimated
slightly deviates from zero, the optimal state is neither the squeezed
vacuum state nor the photon-subtracted squeezed vacuum state, but the
photon-added squeezed vacuum state when they carry many photons. Finally, we
show that the quantum Cram\'{e}r-Rao bound can be reached by parity
detection.

\begin{description}
\item[PACS number(s)] 42.50.Dv,03.65.Ta
\end{description}
\end{abstract}

\maketitle

\section{Introduction}

Phase estimation and optimal interferometry play a significant role for many
precision measurement applications. For a Mach-Zehnder interferometer (MZI),
when only coherent light is injected into one input port of the first beam
splitter, the other input port is by default the vacuum of light, the
sensitivity of the phase estimation is limited by the standard quantum noise
limit (SNL), i.e., $\Delta \phi \propto 1/\sqrt{\bar{N}}$, where $\bar{N}$
the average photon number in the input beam \cite{1,2}. When using a
nonclassical input state, for example, a coherent light and a squeezed
vacuum light are injected into the two input ports of a MZI, Caves \cite{3}
find that the sensitivity of phase estimation below the SNL. Since then, in
order to go beyond the SNL, many highly nonclassical states are employed to
reduce the phase uncertainty \cite{3,4,5,6,7,8,9,10,11,12,13,14,15}, and to
approach $\Delta \phi \propto 1/\bar{N}$, the so-called Heisenberg limit
(HL) \cite{16,17}.

In general, the sensitivity of phase estimation within an interferometer
crucially depends on input states as well as detection schemes. Coherent
states and squeezed vacuum states as well as Fock states are useful for
metrology under the present experimental technology. For example, via the
analysis of the quantum Fisher information, for a MZI with a Fock state in
one input port and an arbitrary state with the same total average photon
number in the other input, same phase uncertainties will be achieved and can
approach to the HL \cite{a2}. Following the theoretical work in Ref. \cite%
{a2}, we analytically prove that the quantum Cramer-Rao bound can be reached
via the parity detection in the limit $\varphi \rightarrow 0$ \cite{a3}. By
Bayesian analysis of the photon number statistics of the output state in a
MZI, Pezz\'{e} and Smerzi show that the HL sensitivity of phase estimation
can be achieved when the coherent light and squeezed vacuum light are mixed
in roughly equal intensities \cite{18}. As a simple alternative to the
detection scheme of Ref.\cite{18}, parity detection for a MZI with coherent
and squeezed vacuum light has bee also investigated \cite{19}. A parity
measurement simply measures the even or odd number of photons in the output
mode \cite{r1}.

Recently, Lang and Caves have considered the question: given that one input
of an interferometer is entered by coherent light, what is the best state to
inject the other input port for achieving high-sensitivity phase-shift
measurements within a constraint on the average photon number that the state
can carry? The answer, they find, is the squeezed vacuum state (SVS) \cite%
{20}. As just pointed out in the conclusion in Ref. \cite{20}, it needs to
further investigate whether the SVS is the optimal state when the squeezing
light carries many photons, even carries as many or more photons than the
coherent input. On the other hand, when the phase shift to be estimated
slightly deviates from zero, whether the SVS with many photons is also the
optimal state? For fixed initial squeezing parameter, Birrittella and Gerry
claim that the corresponding sensitivity of the phase estimation can be
increased via parity detection when the mixing of coherent states and
photon-subtracted SVS (PSSVS) as input states of the MZI \cite{21}. However,
within a constraint on the total average photon number, whether the PSSVS
can indeed improve the phase sensitivity?

In this work, we will answer the above questions by investigating the
interferometry performed by mixing a coherent state with non-Gaussian
squeezed states, such as a photon-added SVS (PASVS) and the PSSVS. At
present, the best experimentally realized non-Gaussian squeezed state in
quantum optics is the photon-subtracted squeezed states \cite{r2,r3}, and
photon subtraction can be implemented by a beam splitter with high
transmissivity \cite{r2}. For the photon addition operation, Agarwal and
Tara \cite{r3} first theoretically studied the nonclassical properties of a
photon-added coherent state. In 2004, the photon addition operation was
successfully demonstrated experimentally via a non-degenerate parametric
amplifier with small coupling strength \cite{r5}. We also compared the phase
sensitivity with another quantum limit, the quantum Cram\'{e}r-Rao bound
\cite{a4}, which sets the ultimate limit for a set of probabilities that
originated from measurements on a quantum system.

The organization of this paper is as follows. In Sec. II, we make a brief
review about both the PASVS and the PSSVS. And then we describes the
propagation of a two-mode light, initially in the product state of PASVS and
coherent light, through the MZI. Section III focuses on the parity detection
scheme and provides the phase sensitivity. We will show that for giving a
constraint on the average photon numbers of PASVS/PSSVS/SVS, almost the same
phase sensitivity will be achieved by parity detection with such kinds of
input states. Especially, when the phase shift slightly deviates from zero,
the optimal state is neither the SVS nor the PSSVS, but the PASVS when they
carry many photons. In Sec. IV, we prove that the quantum Cramer-Rao bound
can be reached via the parity detection in the limit $\varphi \rightarrow 0$.

\section{Resulted output state of the input fields through the interferometer%
}

Previously, Birrittella and Gerry \cite{21} studied the prospect of
parity-based interferometry with mixing a PSSVS and a coherent state. In
contrast to that scheme, we mainly investigate the scheme when a PASVS and a
coherent state ($\left \vert z\right \rangle $ with the amplitude parameter $%
z=\left \vert z\right \vert e^{\theta }$) are considered as the input state of
a MZI. For our purposes, we first provide a brief review of both the PASVS
and the PSSVS which is a kind of non-Gaussian squeezed vacuum states. And
then, we derive the resulted output state when a PASVS and a coherent state
are injected into a balanced MZI

\subsection{Photon-added and photon-subtracted squeezed vacuum states}

The SVS is a Gaussian state, which is defined as \cite{22}%
\begin{equation}
\left \vert r\right \rangle =S\left( r\right) \left \vert 0\right \rangle =%
\mathrm{sech}^{1/2}r\exp \left( -\frac{1}{2}b^{\dag 2}\tanh r\right) \left
\vert 0\right \rangle ,  \label{1}
\end{equation}%
where $S\left( r\right) =\exp \left[ r\left( b^{2}-b^{\dag 2}\right) /2%
\right] $ with the squeezing parameter $r$. By repeatedly operating the
photon addition operator $b^{\dag }$ on a SVS, one obtain the normalized
PASVS which is defined as
\begin{equation}
\left \vert r,k\right \rangle =\frac{1}{\sqrt{N_{k}}}b^{\dag k}S\left(
r\right) \left \vert 0\right \rangle ,  \label{2}
\end{equation}%
where $N_{k}$ is the corresponding normalization factor \cite{23}%
\begin{eqnarray}
N_{k} &=&\frac{\partial ^{2k}}{\partial t^{k}\partial \tau ^{k}}\exp \left[ -%
\frac{\sinh 2r}{4}\left( t^{2}+\tau ^{2}\right) +t\tau \cosh ^{2}r\right]
|_{t,\tau =0}  \notag \\
&=&k!\cosh ^{k}rP_{k}\left( \cosh r\right) .  \label{3}
\end{eqnarray}%
The average photon number of the PASVS is given by%
\begin{equation}
\bar{n}_{a}=\left \langle b^{\dag }b\right \rangle _{\text{PASVS}}=\frac{%
N_{k+1}}{N_{k}}-1.  \label{5}
\end{equation}

Different from that in Ref.\cite{21}, we adopt the expression of the PSSVS
as following
\begin{equation}
\left \vert r,-l\right \rangle =\frac{1}{\sqrt{C_{l}}}b^{l}S\left( r\right)
\left \vert 0\right \rangle ,  \label{6}
\end{equation}%
where $b$ is the photon subtraction operator, and the normalization factor
is \cite{24}%
\begin{eqnarray}
C_{l} &=&\frac{\partial ^{2l}}{\partial t^{l}\partial \tau ^{l}}\exp \left[ -%
\frac{\sinh 2r}{4}\left( t^{2}+\tau ^{2}\right) +t\tau \sinh ^{2}r\right]
|_{t,\tau =0}  \notag \\
&=&l!\left( -i\sinh r\right) ^{l}P_{l}\left( i\sinh r\right) .  \label{7}
\end{eqnarray}%
The average photon number of the PSSVS is given by%
\begin{equation}
\bar{n}_{s}=\left \langle b^{\dag }b\right \rangle _{\text{PSSVS}}=\frac{%
C_{k+1}}{C_{k}}.  \label{8}
\end{equation}%
Particularly, when $k=l$, the single-photon PASVS and the single-photon
PSSVS are the same non-Gaussian squeezed state \cite{25}.

For the quantum metrology, the average photon number of input states is an
important factor. Actually, both photon addition and subtraction can
increase the average photon number of the state as shown in Fig. 1. Clearly,
it is always better to perform addition rather than subtraction in order to
increase the average photon number for given the initial squeezing. And
hence, we expect that the PASVS can offer an improved phase resolution over
both the PSSVS-coherent input state and SVS-coherent input state for a given
initial squeezing parameter of the SVS. The fact that the photon subtraction
can increase the average photon number of the states with super-Poissonian
statistics was explained in detail very recently by Stephen et al \cite{r6}
and references involved in that. Indeed, the distribution for the
single-mode SVS, as is well known, is super-Poissonian \cite{r8}.

\begin{figure}[tbph]
\centering \includegraphics[width=8cm]{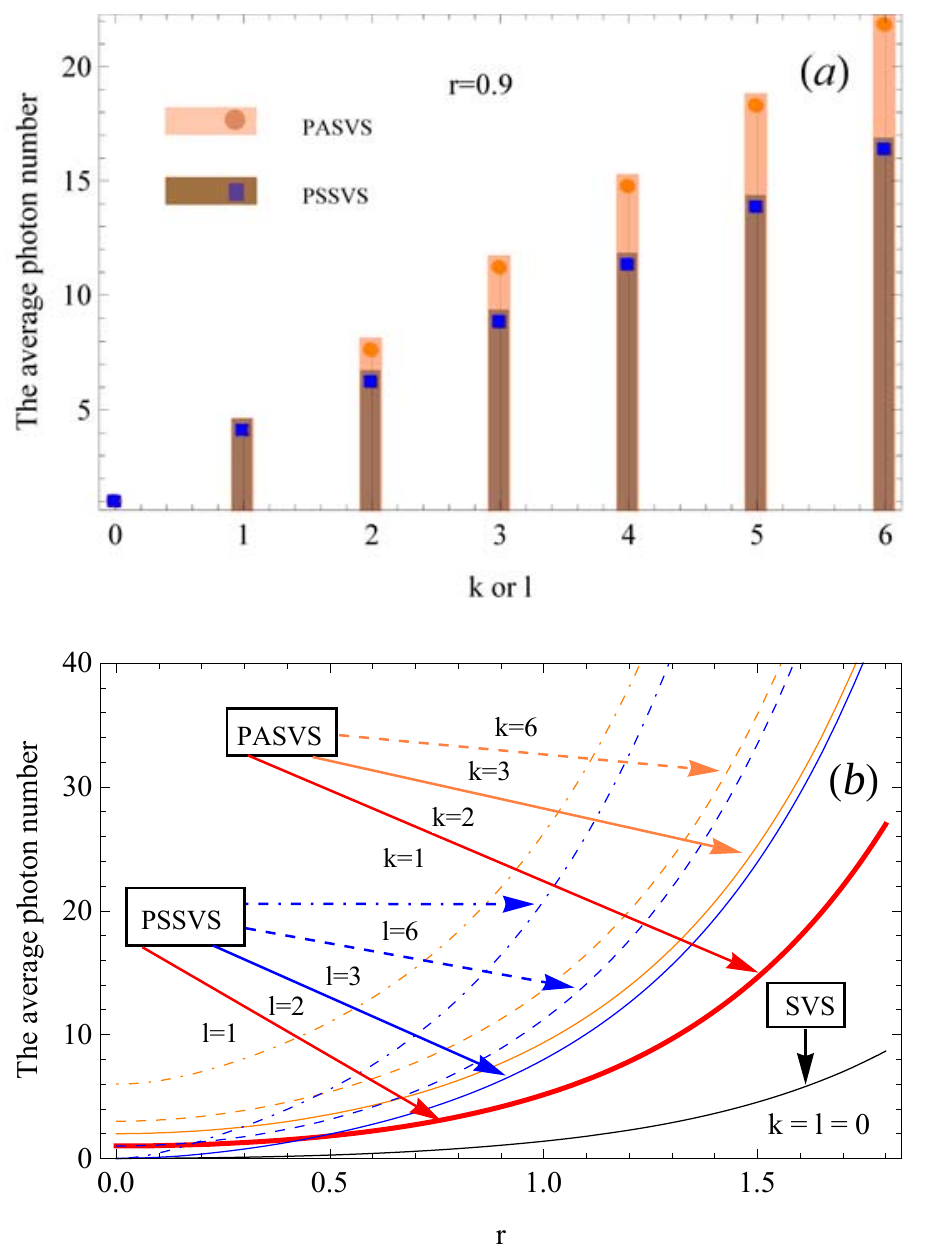}
\caption{Plots of the average photon number of both the PASVS and the PSSVS,
as well as the SVS, respectively. ($a$). The average photon number for a
given SVS (fixed squeezing parameter $r$) as a function of the number $k$ or
$l$; ($b$) The average photon number as a function of the squeezing
parameter $r$ for some values of $k$ and $l$.}
\end{figure}

\subsection{Resulted output state of the PASVS through the interferometer}

The balanced MZI considered here is mainly composed of two 50:50 beam
splitters and two phase shifters. Generaly, the first beam splitter BS1 is
described by the transformation $U_{\text{BS1}}=\exp \left[ -i\pi \left(
a^{\dagger }b+ab^{\dagger }\right) /4\right] .$And the operator
representation of the second beam splitter BS2 is taken as $U_{\text{BS2}%
}=\exp \left[ i\pi \left( a^{\dagger }b+ab^{\dagger }\right) /4\right] $.
The operator $U\left( \varphi \right) =\exp \left[ i\varphi \left(
a^{\dagger }a-b^{\dagger }b\right) /2\right] $ represents the two phase
shifters, the angle $\varphi $ being the phase shift between the two arms to
be estimated. The unitary transformation associated with such balanced MZI
can be written as \cite{14}%
\begin{equation}
U\left( \varphi \right) =e^{i\frac{\pi }{2}J_{1}}e^{-i\varphi J_{3}}e^{-i%
\frac{\pi }{2}J_{1}}=e^{-i\varphi J_{2}},  \label{9}
\end{equation}%
where these operators consisted of two sets of Bose operator
\begin{eqnarray}
J_{1} &=&\frac{1}{2}\left( a^{\dagger }b+ab^{\dagger }\right) ,\text{ }J_{2}=%
\frac{1}{2i}\left( a^{\dagger }b-ab^{\dagger }\right) ,  \notag \\
\text{\ }J_{3} &=&\frac{1}{2}\left( a^{\dagger }a-b^{\dagger }b\right) ,
\label{10}
\end{eqnarray}%
are the angular momentum operators in the well-known Schwinger
representation \cite{r9}. They satisfy the commutation relation $\left[
J_{i},J_{j}\right] =i\epsilon _{ijk}J_{k}$ ($i,j,k=1,2,3$), and commute with
the Casimir operator $J_{0}=\frac{1}{2}\left( a^{\dagger }a+b^{\dagger
}b\right) $, i.e., $\left[ J_{0},J_{i}\right] =0$. Propagation of the input
fields (pure states) through these elements, the resulted output state can
be written as
\begin{equation}
\left \vert \text{out}\right \rangle _{\text{MZI}}=e^{-i\varphi J_{2}}\left
\vert \psi \right \rangle _{\text{in}}.  \label{11}
\end{equation}%
For the balanced MZI, applying the following transformation relations%
\begin{eqnarray}
e^{-i\varphi J_{2}}a^{\dagger }e^{i\varphi J_{2}} &=&a^{\dagger }\cos \frac{%
\varphi }{2}+b^{\dagger }\sin \frac{\varphi }{2},  \notag \\
e^{-i\varphi J_{2}}b^{\dagger }e^{i\varphi J_{2}} &=&b^{\dagger }\cos \frac{%
\varphi }{2}-a^{\dagger }\sin \frac{\varphi }{2}  \label{12}
\end{eqnarray}%
and the relation $e^{-i\varphi J_{2}}\left \vert 0\right \rangle
_{a}\left
\vert 0\right \rangle _{b}=\left \vert 0\right \rangle
_{a}\left
\vert 0\right \rangle _{b}$, in principle, one can obtain the
explicit form of the output state the MZI.

For the convenience of the later calculation, we rewrite the PASVS $%
\left \vert \psi _{r,k}\right \rangle _{b}$ in the basis of the coherent state
as follows%
\begin{equation}
\left \vert \psi _{r,k}\right \rangle _{b}=\frac{\mathrm{sech}^{1/2}r}{\sqrt{%
N_{k}}}\frac{d^{k}}{dg^{k}}\int \frac{d^{2}\alpha }{\pi }e^{-\frac{%
\left \vert \alpha \right \vert ^{2}}{2}+g\alpha ^{\ast }-\frac{\tanh r}{2}%
\alpha ^{\ast 2}}\left \vert \alpha \right \rangle |_{g=0},  \label{13}
\end{equation}%
where $\left \vert \alpha \right \rangle =\exp \left[ -\left \vert \alpha
\right \vert ^{2}/2+\alpha b^{\dag }\right] \left \vert 0\right \rangle $ is a
coherent state. When the product state $\left \vert \psi \right \rangle _{%
\text{in}}=\left \vert z\right \rangle _{a}\otimes \left \vert \psi
_{r,k}\right \rangle _{b}$ is injected into the MZI, the resulted output
state can be written as%
\begin{eqnarray}
\left \vert \psi \right \rangle _{\text{out}} &=&\frac{\mathrm{sech}^{1/2}r}{%
\sqrt{N_{k}}}\frac{d^{k}}{dg^{k}}\int \frac{d^{2}\alpha }{\pi }e^{-\frac{%
\left \vert z\right \vert ^{2}}{2}-\left \vert \alpha \right \vert ^{2}+g\alpha
^{\ast }-\frac{\tanh r}{2}\alpha ^{\ast 2}}  \notag \\
&&e^{\left( \alpha \sin \frac{\varphi }{2}+z\cos \frac{\varphi }{2}\right)
a^{\dagger }+\allowbreak \left( \allowbreak \alpha \cos \frac{\varphi }{2}%
-z\sin \frac{\varphi }{2}\right) b^{\dagger }}\left \vert 0,0\right \rangle
|_{g=0},  \label{15}
\end{eqnarray}%
which is the state of light at the output of the MZI. In the following
section, we shall mainly present parity measurement scheme with calculations
of the expected value of the parity operator and the corresponding phase
sensitivity.

\section{Phase estimation with parity detection}

There are several detection methods for extracting phase information from
the output states of the MZI, e.g., intensity detection, homodyne detection,
and parity detection \cite{r1}. Parity detection simply measures the even or
odd number of photons in the output mode. In addition, as shown in Ref. \cite%
{19}, the parity detection saturates the quantum Cram\'{e}r-Rao bound and in
turn provides the HL phase sensitivity when the SVS-coherent state are mixed
in equal proportions. In experiments, the parity detection using a
photon-number resolving detector with coherent states \cite{r10} has also
been demonstrated. Here, we use the parity detection too.

\subsection{The parity detection with the PASVS-coherent state}

For the detailed discussion of the parity detection in quantum optimal
metrology, one can review that in Ref.\cite{r1}. Actually, the parity
detection is to obtain the expectation value of the parity operator in the
output state of the MZI. In order to calculate conveniently, using the
operator identity $\exp \left( \lambda a^{\dag }a\right) =\colon \exp \left[
\left( e^{\lambda }-1\right) a^{\dag }a\right] \colon $, we rewrite the
parity operator as follows%
\begin{equation}
\Pi _{b}=\left( -1\right) ^{b^{\dag }b}=e^{i\pi b^{\dag }b}=\colon
e^{-2b^{\dag }b}\colon =\int \frac{d^{2}\gamma }{\pi }\left \vert \gamma
\right \rangle \left \langle -\gamma \right \vert ,  \label{16}
\end{equation}%
where $\left \vert \gamma \right \rangle \,$\ is a coherent state, and $\colon
\colon $ denotes the normally ordered form of Bose operators. Here, we
consider performing parity detection on just one of the output modes, for
instance, the $b$ mode. The parity operator on an output mode $b$ is
described by $\Pi _{B}=\left( -1\right) ^{b^{\dag }b}$, then the expectation
value of the parity operator is
\begin{equation}
\left \langle \Pi _{b}\left( \varphi \right) \right \rangle =_{\text{out}%
}\left \langle \psi \right \vert \int \frac{d^{2}\gamma }{\pi }\left \vert
\gamma \right \rangle \left \langle -\gamma \right \vert \left \vert \psi
\right \rangle _{\text{out}}.  \label{17}
\end{equation}%
Now, we consider the corresponding the expectation value of the parity
operator in the\ output state when the PASVS-coherent state is injected into
the MZI. Then substituting Eqs. (\ref{15}) and (\ref{16}) into Eq. (\ref{17}%
), and applying the integral formula
\begin{equation}
\int \frac{d^{2}z}{\pi }e^{\zeta \left \vert z\right \vert ^{2}+\xi z+\eta
z^{\ast }+fz^{2}+gz^{\ast 2}}=\frac{1}{\sqrt{\zeta ^{2}-4fg}}e^{\frac{-\zeta
\xi \eta +\xi ^{2}g+\eta ^{2}f}{\zeta ^{2}-4fg}}  \label{18}
\end{equation}%
whose convergent condition is Re$\left( \xi \pm f\pm g\right) <0\ $ and Re$%
\left( \frac{\zeta ^{2}-4fg}{\xi \pm f\pm g}\right) <0$, after directly
calculation, we obtain%
\begin{eqnarray}
\left \langle \Pi _{b}\left( \varphi \right) \right \rangle _{\text{PA}} &=&%
\frac{\left \langle \Pi _{b}\left( \varphi \right) \right \rangle _{0}}{N_{k}}%
\frac{\partial ^{2k}}{\partial h^{k}\partial g^{k}}\exp \left[ -\frac{\cosh
^{2}r\cos \varphi }{1+\sin ^{2}\varphi \sinh ^{2}r}gh\right.   \notag \\
&&\left. +\frac{4z\cosh ^{2}r\sin \varphi +z^{\ast }\sinh 2r\sin 2\varphi }{%
4\left( 1+\sin ^{2}\varphi \sinh ^{2}r\right) }h\right.   \notag \\
&&\left. +\frac{4z^{\ast }\cosh ^{2}r\sin \varphi +z\sinh 2r\sin 2\varphi }{%
4\left( 1+\sin ^{2}\varphi \sinh ^{2}r\right) }g\right.   \notag \\
&&\left. -\frac{\sinh 2r\cos ^{2}\varphi }{4\left( 1+\sinh ^{2}r\sin
^{2}\varphi \right) }\left( h^{2}+\allowbreak g^{2}\right) \right] |_{h=g=0},
\label{19}
\end{eqnarray}%
where $\left \langle \Pi _{b}\right \rangle _{0}$ is the corresponding
expectation value of the parity operator for the input state with
SVS-coherent state \cite{19},%
\begin{equation}
\left \langle \Pi _{b}\left( \varphi \right) \right \rangle _{0}=\frac{e^{%
\frac{2\left( \cos \varphi -1-\sinh ^{2}r\sin ^{2}\varphi \right) \left \vert
z\right \vert ^{2}-\sinh 2r\sin ^{2}\varphi \mathrm{Re}\left( z^{2}\right) }{%
2\left( 1+\sinh ^{2}r\sin ^{2}\varphi \right) }}}{\sqrt{1+\sinh ^{2}r\sin
^{2}\varphi }}.  \label{20}
\end{equation}

Because the goal of the interferometry is to estimate very small phase
changes in quantum metrology, it may be interesting to expand Eq. (\ref{19})
in the Taylor series around $\varphi =0$. To write out the explicit Taylor
expansion of Eq. (\ref{19}) for general $k$\ in the limit $\varphi
\rightarrow 0$\ is a difficult task. However, for small $k$, the explicit
Taylor expansions can be accessible. When $k=0$, the Taylor expansions of $%
\left \langle \Pi _{b}\left( \varphi \right) \right \rangle _{0}$\ is

\begin{eqnarray}
&&\left \langle \Pi _{b}\left( \varphi \rightarrow 0\right) \right \rangle
_{0}  \notag \\
&=&\left[ 1-\bar{n}_{z}\cos \theta \sqrt{\bar{n}_{a}^{2}+\bar{n}_{a}+1}%
\varphi ^{2}\right.  \notag \\
&&\left. -\frac{2\bar{n}_{z}\bar{n}_{a}+\bar{n}_{z}+\bar{n}_{a}}{2}\varphi
^{2}+O\left( \varphi ^{4}\right) \right] ,  \label{21}
\end{eqnarray}%
where $\bar{n}_{z}=\left \vert z\right \vert ^{2}$\ ($z=\left \vert
z\right
\vert e^{i\theta }$) the average photon number of the coherent
state, $\bar{n}_{a}=\left( N_{k+1}/N_{k}-1\right) |_{k=0}=\sinh ^{2}r$ the
average photon number of the SVS. While $k=1,2$, we can derive the
analytical Taylor expansions of $\left \langle \Pi _{b}\left( \varphi
\right) \right \rangle _{\text{PA}}$ as well, i.e.,
\begin{eqnarray}
&&\left \langle \Pi _{b}\left( \varphi \rightarrow 0\right) \right \rangle _{%
\text{PA}}|_{k=1}  \notag \\
&=&\left( -1\right) \left[ 1-\bar{n}_{z}\cos \theta \sqrt{\bar{n}_{a}^{2}+%
\bar{n}_{a}-2}\varphi ^{2}\right.  \notag \\
&&\left. -\frac{2\bar{n}_{z}\bar{n}_{a}+\bar{n}_{z}+\bar{n}_{a}}{2}\varphi
^{2}+O\left( \varphi ^{4}\right) \right] ,  \label{22}
\end{eqnarray}%
and%
\begin{eqnarray}
&&\left \langle \Pi _{b}\left( \varphi \rightarrow 0\right) \right \rangle _{%
\text{PA}}|_{k=2}  \notag \\
&=&\left[ 1-\bar{n}_{z}\cos \theta \sqrt{\bar{n}_{a}^{2}+\bar{n}_{a}-\frac{%
8\left( \sqrt{1+12\bar{n}_{a}}+1\right) ^{2}}{3\left( \sqrt{1+12\bar{n}_{a}}%
-1\right) ^{2}}}\varphi ^{2}\right.  \notag \\
&&\left. -\frac{2\bar{n}_{z}\bar{n}_{a}+\bar{n}_{z}+\bar{n}_{a}}{2}\varphi
^{2}+O\left( \varphi ^{4}\right) \right] ,  \label{23}
\end{eqnarray}%
where $\bar{n}_{a}=N_{k+1}/N_{k}-1$\ the average photon number of the PASVS.
In order to obtain the good phase uncertainty, in the following we set $%
\theta =0$ (the phase of the coherent state) \cite{19}. These analytical
Taylor expansions of $\left \langle \Pi _{b}\left( \varphi \right)
\right
\rangle _{\text{PA}}$\ mentioned above is useful for the following
discussions about the parity detection.

Different from that calculation method in Ref.\cite{21}, for the normalized
PSSVS, using our method, we can also obtain the corresponding expectation
value of the parity operator for the input state with PSSVS-coherent states

\begin{eqnarray}
\left \langle \Pi _{b}\left( \varphi \right) \right \rangle _{\text{PS}} &=&%
\frac{\left \langle \Pi _{b}\left( \varphi \right) \right \rangle _{0}}{C_{l}%
}\frac{\partial ^{2l}}{\partial h^{l}\partial g^{l}}\exp \left[ -\frac{\sinh
^{2}r\cos \varphi }{1+\sin ^{2}\varphi \sinh ^{2}r}gh\right.  \notag \\
&&\left. -\frac{z\sinh ^{2}r\sin 2\varphi +z^{\ast }\sinh 2r\sin \varphi }{%
2\left( 1+\sin ^{2}\varphi \sinh ^{2}r\right) }g\right.  \notag \\
&&\left. -\frac{z^{\ast }\sinh ^{2}r\sin 2\varphi +z\sinh 2r\sin \varphi }{%
2\left( 1+\sin ^{2}\varphi \sinh ^{2}r\right) }h\right.  \notag \\
&&\left. -\frac{\sinh r\cosh r}{2\left( 1+\sin ^{2}\varphi \sinh
^{2}r\right) }\left( h^{2}+\allowbreak g^{2}\right) \right] |_{h=g=0},
\label{27}
\end{eqnarray}%
Although Eq. (\ref{27}) is different from Eq. (16) in Ref.\cite{21}, one can
prove that two results are completely identical by numerical method. In
addition, when \thinspace $k=l=1,$ one can obtain $\left \langle \Pi
_{b}\left( \varphi \right) \right \rangle _{\text{PS}}|_{l=1}=\left \langle
\Pi _{b}\left( \varphi \right) \right \rangle _{\text{PA}}|_{k=1}$, this is
because one photon-subtracted SVS is identical to one photon-added SVS \cite%
{25}. When $l=2$, the Taylor expansion of $\left \langle \Pi _{b}\left(
\varphi \right) \right \rangle _{\text{PS}}$\ in the limit $\varphi
\rightarrow 0$ is%
\begin{eqnarray}
&&\left \langle \Pi _{b}\left( \varphi \right) \right \rangle _{\text{PS}%
}|_{l=2}  \notag \\
&=&\left[ 1-\bar{n}_{z}\cos \theta \sqrt{\bar{n}_{s}^{2}+\bar{n}_{s}-\frac{%
8\left( \sqrt{1+12\bar{n}_{s}}-1\right) ^{2}}{3\left( \sqrt{1+12\bar{n}_{s}}%
+1\right) ^{2}}}\varphi ^{2}\right.  \notag \\
&&\left. -\frac{2\bar{n}_{z}\bar{n}_{s}+\bar{n}_{z}+\bar{n}_{s}}{2}\varphi
^{2}+O\left( \varphi ^{4}\right) \right]  \label{29}
\end{eqnarray}%
where $\bar{n}_{s}=C_{l+1}/C_{l}$\ the average photon number of the PSSVS.

By Eqs. (\ref{19}) and (\ref{27}), we can investigate the expectation value
of the parity operator as function of the phase shift $\varphi $. Given $%
r=0.3$ and $z=2$, in Fig. 2($a$) we draft this expectation value against $%
\varphi $ for $k=0,1,2,3$ and $l=0,1,2,3$. One can see that, for the central
peak or trough of the $\left \langle \Pi _{b}\left( \varphi \right)
\right
\rangle $, the narrowness of the maxima or minima at $\varphi =0$
increases as $k$ and $l$ increase. In addition, we see that the central peak
or trough of the $\left \langle \Pi _{b}\right \rangle _{\text{PA}}$\ of the
PASVS-coherent input states is narrower than that of the PSSVS-coherent
input states when \thinspace $k=l>1.$This result indicates that the photon
addition can enhance supper-resolution better than photon subtraction for
given initial coherent state amplitude and squeezing parameters. However, if
given the same average photon number ($\bar{n}_{a}=\bar{n}_{s}$)\ of both
the PASVS and the PSSVS, the distributions of these central peaks or troughs
of the $\left \langle \Pi _{b}\left( \varphi \right) \right \rangle $\ near $%
\varphi =0$\ are almost identical as shown in Fig. 2($b$), which is also
consistent with Eqs. (\ref{21}-\ref{23}) and (\ref{29}).

\begin{figure}[tbph]
\centering \includegraphics[width=8cm]{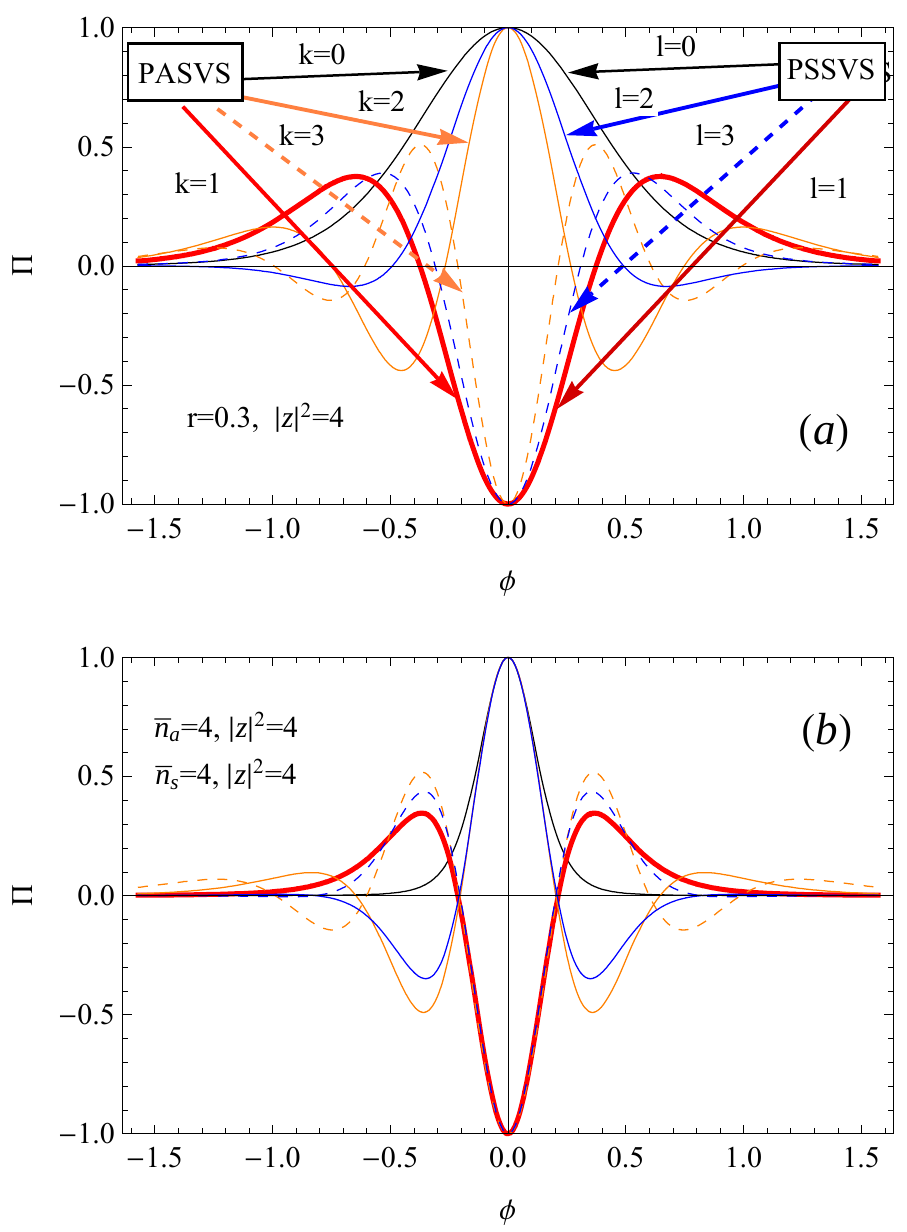}
\caption{The expectation value of the parity operator versus the phase shift
$\protect \varphi $ for different PASVS-coherent and PSSVS-coherent input
states of the MZI, respectively.}
\end{figure}

\subsection{Phase sensitivity via the error propagation method}

The phase uncertainty $\Delta \varphi $ is the main aspect of quantum
optimal interferometry. The smallest phase uncertainty $\Delta \varphi $ is
the characteristic of the most sensitive measure. The smaller the value of $%
\Delta \Pi _{b}$ is, the higher the phase sensitivity is. From the error
propagation method, the phase uncertainty $\Delta \varphi $ of an
interferometer can be determined as%
\begin{equation}
\Delta \varphi =\frac{\sqrt{1-\left \langle \Pi _{b}\left( \varphi \right)
\right \rangle ^{2}}}{\left \vert \partial \left \langle \Pi _{b}\left( \varphi
\right) \right \rangle /\partial \varphi \right \vert }  \label{30}
\end{equation}%
where we have used $\Delta \Pi _{b}=\sqrt{\left \langle \Pi _{b}^{2}\left(
\varphi \right) \right \rangle -\left \langle \Pi _{b}\left( \varphi \right)
\right \rangle ^{2}}$ and the fact that $\left \langle \Pi _{b}^{2}\left(
\varphi \right) \right \rangle =1$. The phase sensitivity with parity
detection for the interferometry with PASVS-coherent state is found to be
best at $\varphi =0$. Although it is difficult to write out the general
explicit form of Eq. (\ref{30}) when the PASVS-coherent state with the
general value of $k$ is considered as an interferometer state, based on Eqs.
(\ref{21}-\ref{23}), the explicitly forms of $\Delta \varphi $ for small $k$
in the limit of $\varphi \rightarrow 0$ can be easily obtained. For example,
according to Eq. (\ref{21}), the explicit form of Eq. (\ref{30}) in the case
of $\varphi \rightarrow 0$ can be directly obtained \cite{19},
\begin{equation}
\left( \Delta \varphi \right) _{\text{PA}}|_{k=0}=\frac{1}{\sqrt{2\bar{n}_{z}%
\sqrt{\bar{n}_{a}^{2}+\bar{n}_{a}+1}+2\bar{n}_{z}\bar{n}_{a}+\bar{n}_{z}+%
\bar{n}_{a}}},  \label{31}
\end{equation}%
where they have set $\theta =0$ (the phase of the coherent state) in order
to obtain the good phase uncertainty. In addition, based on our Eqs. (\ref%
{22}) and (\ref{23}), we can also obtain the explicit form of Eq. (\ref{30})
for $k=1,2$, respectively,

\begin{equation}
\left( \Delta \varphi \right) _{\text{PA}}|_{k=1}=\frac{1}{\sqrt{2\bar{n}_{z}%
\sqrt{\bar{n}_{a}^{2}+\bar{n}_{a}-2}+2\bar{n}_{z}\bar{n}_{a}+\bar{n}_{z}+%
\bar{n}_{a}}},  \label{32}
\end{equation}%
and
\begin{eqnarray}
&&\left( \Delta \varphi \right) _{\text{PA}}|_{k=2}  \notag \\
&=&\frac{1}{\sqrt{2\bar{n}_{z}\sqrt{\bar{n}_{a}^{2}+\bar{n}_{a}-\frac{%
8\left( \sqrt{1+12\bar{n}_{a}}+1\right) ^{2}}{3\left( \sqrt{1+12\bar{n}_{a}}%
-1\right) ^{2}}}+2\bar{n}_{z}\bar{n}_{a}+\bar{n}_{z}+\bar{n}_{a}}}.
\label{33}
\end{eqnarray}%
For the item $\frac{8\left( \sqrt{1+12\bar{n}_{a}}+1\right) ^{2}}{3\left(
\sqrt{1+12\bar{n}_{a}}-1\right) }$ in Eq. (\ref{33}), one can easily obtain $%
\frac{8\left( \sqrt{1+12\bar{n}_{a}}+1\right) ^{2}}{3\left( \sqrt{1+12\bar{n}%
_{a}}-1\right) }|_{k=2}=\frac{24\cosh ^{4}r}{(3\cosh ^{2}r-1)^{2}}$, and
prove that the inequality $\frac{8}{3}\leq \frac{8\left( \sqrt{1+12\bar{n}%
_{a}}+1\right) ^{2}}{3\left( \sqrt{1+12\bar{n}_{a}}-1\right) ^{2}}\leq $ $6$
for any values of the squeezing parameter $r$ is satisfied. On the other
hand, when the PSSVS-coherent state with $l=2$ is considered as the
interferometer state, according to Eq. (\ref{29}), we obtain the
corresponding phase uncertainty in the case of $\varphi \rightarrow 0$ as%
\begin{eqnarray}
&&\left( \Delta \varphi \right) _{\text{PS}}|_{l=2}  \notag \\
&=&\frac{1}{\sqrt{2\bar{n}_{z}\sqrt{\bar{n}_{s}^{2}+\bar{n}_{s}-\frac{%
8\left( \sqrt{1+12\bar{n}_{s}}-1\right) ^{2}}{3\left( \sqrt{1+12\bar{n}_{s}}%
+1\right) ^{2}}}+2\bar{n}_{z}\bar{n}_{s}+\bar{n}_{z}+\bar{n}_{s}}}
\label{34}
\end{eqnarray}%
where we have set $\theta =0$ as well. For the item $\frac{8\left( \sqrt{1+12%
\bar{n}_{s}}-1\right) ^{2}}{3\left( \sqrt{1+12\bar{n}_{s}}+1\right) ^{2}}%
|_{l=2}=\frac{24\sinh ^{4}r}{(3\sinh ^{2}r+1)^{2}}$ in Eq. (\ref{34}), one
can easily prove that the inequality $0\leq 24\sinh ^{4}r(3\sinh
^{2}r+1)^{-2}\leq $ $\frac{8}{3}$ is satisfied.\textbf{\ }

Obviously, Eqs. (\ref{31}-\ref{34}) indicate that, within constraints on the
total average photon number ($\bar{N}=\bar{n}_{a}+\bar{n}_{z}$ or $\bar{N}=%
\bar{n}_{s}+\bar{n}_{z}$) and the same ratios of $\bar{n}_{a}/\bar{n}_{z}$
or $\bar{n}_{s}/\bar{n}_{z}$, the best phase uncertainty is obtained by the
SVS when one input port of a MZI is injected by a coherent state. When $k=l=0
$, both PASVS and PSSVS reduce to a SVS. Therefore, our results obtained via
parity detection indeed support Lang and Caves's work that the SVS is the
optimal state for a MZI in the limit of $\varphi \rightarrow 0$\ when one
input is a coherent state \cite{20}.\ On the other hand, given a larger
value of the same average photon numbers of the PASVS, the PSSVS and the
SVS, we can also see from Eqs. (\ref{31}-\ref{34}) that the difference among
these corresponding phase uncertainties is very small. With the increases of
the values of the $k$\ and $l$, we can numerically prove such conclusion is
still true. As pointed in Refs. \cite{18,20}, we can also prove that, in the
case of the $\bar{n}_{a}=\bar{n}_{z}$ (or $\bar{n}_{s}=\bar{n}_{z}$ ), the
best phase sensitivity will be obtained and reach the HL.

However, when the phase shift $\varphi $\ to be estimated slightly deviates
from zero, the SVS with many photons may be not the optimal state for a MZI.
Firstly, we investigate how the phase uncertainty varies with $\varphi $.
Based on Eq. (\ref{30}), we plot the phase uncertainty as a function of the
phase shift $\varphi $ in Fig. 3. Fig. 3($a$) shows that, given somewhat
small average photon numbers $\bar{n}_{a}=$ $\bar{n}_{s}=\bar{n}_{z}=4$, the
MZI with one input port injected by a coherent state, the SVS state is the
optimal state for such MZI. Yet, interferometry with coherent states
contained a few average number of photons is meaningless since such a
measurement cannot be precise in principle. Moreover as pointed in the
conclusion part in Ref. \cite{20}, it may interesting to explore how the
phase sensitivity changes when the SVS carry many photons. Thus, with the
increases of the value of $\bar{n}_{a}$ $($or $\bar{n}_{s})$ and $\bar{n}_{z}
$, we can see from Fig. 3($b$) that the differences among these
corresponding phase uncertainties are indeed very small in the limit of $%
\varphi \rightarrow 0$. While the phase shift $\varphi $\ deviates from
zero, the optimal state is neither the SVS nor the PSSVS, but the PASVS as
shown in Fig. 3($b$). On the other hand, for a given initial squeezing
parameter $r$ or given the same ratio of $\bar{n}_{a}/\bar{n}_{z}$ (and $%
\bar{n}_{s}/\bar{n}_{z}$) of the two input ports of the MZI, the PASVS has
also the better performance in quantum precision measurement than both the
PSSVS and the SVS when the phase shift $\varphi $\ slightly deviates from
zero as shown in Fig. 4.
\begin{figure}[tbph]
\centering \includegraphics[width=8cm]{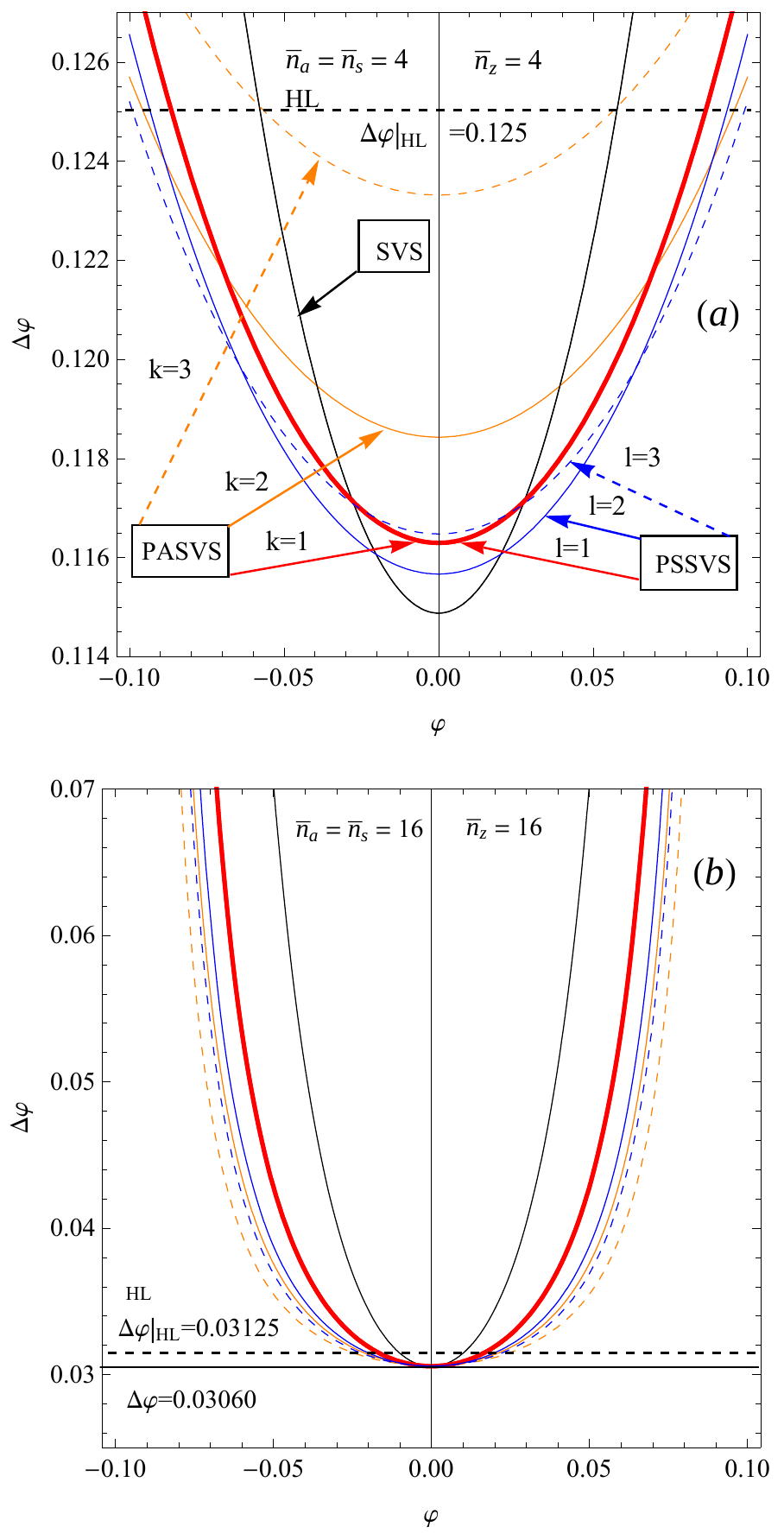}
\caption{The phase uncertainty $\Delta \protect \varphi $ as a function of
the phase shift $\protect \varphi $ when the PASVS-coherent state, the
PSSVS-coherent state and the SVS-coherent state as interferometer states,
respectively. ($a$) $\bar{n}_{a}=$ $\bar{n}_{s}=\bar{n}_{z}=4$; ($b$) $\bar{n%
}_{a}=$ $\bar{n}_{s}=\bar{n}_{z}=16$. The horizontal dashed line denotes the
HL limit.}
\end{figure}
\begin{figure}[tbph]
\centering \includegraphics[width=8cm]{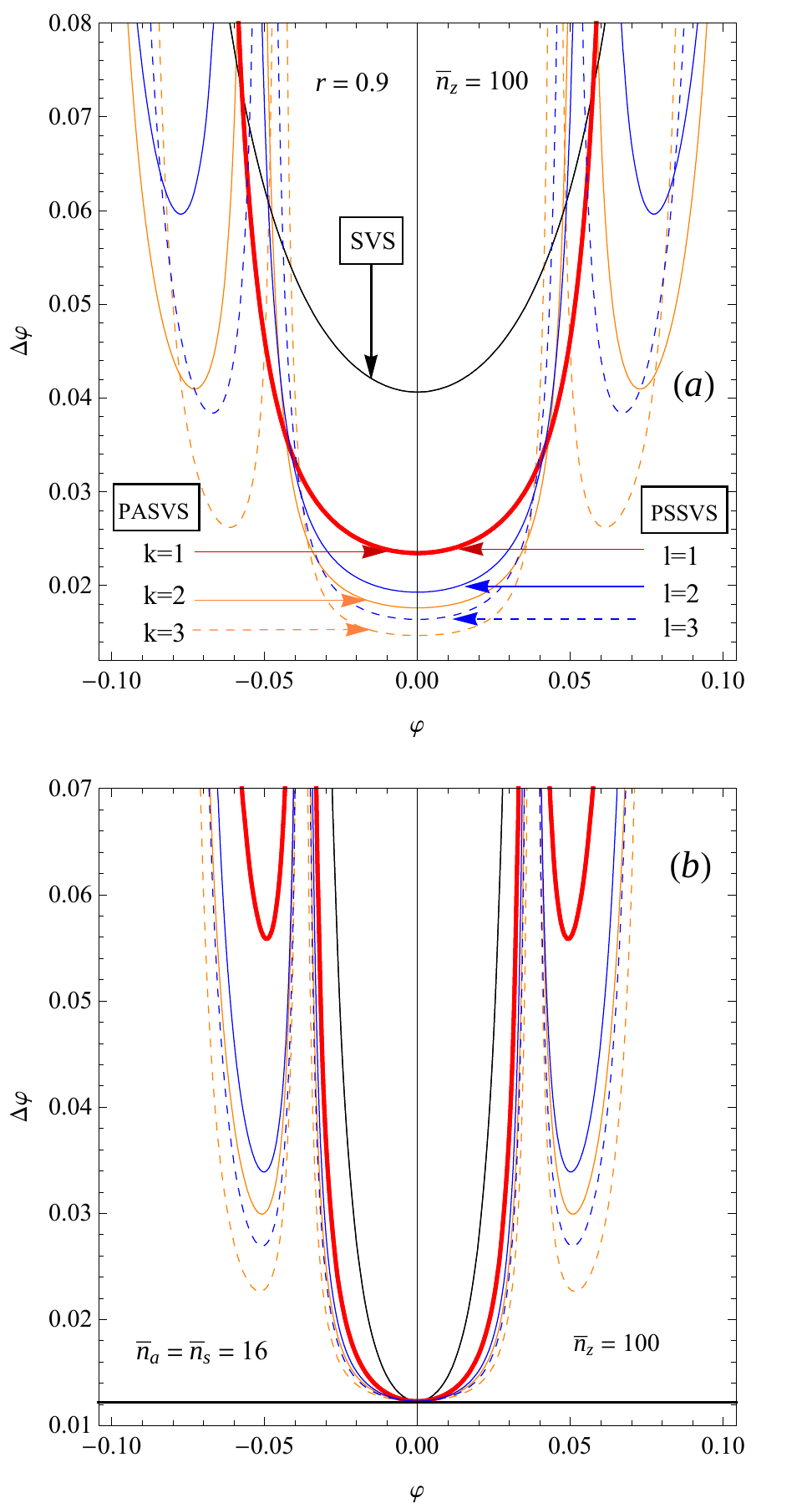}
\caption{The phase uncertainty $\Delta \protect \varphi $ as a function of
the phase shift $\protect \varphi $ when the PASVS-coherent state, the
PSSVS-coherent state and the SVS-coherent state as interferometer states. ($a
$) $r=0.9,\bar{n}_{z}=100$; ($b$) $\bar{n}_{a}=\bar{n}_{s}=16,\bar{n}_{z}=100
$.}
\end{figure}

Secondly, for given a constraint on the total average photon number of the
PASVS, the PSSVS and the SVS, it is important to investigate how the phase
sensitivities change with the total average photon number. In Fig. 5,\ for
the PASVS ($k=0,1,2,3,6)$ and the PSSVS ($l=0,1,2,3,6$), we give some same
values of both $\bar{n}_{a}$ and $\bar{n}_{s}$ and plot the phase
sensitivities versus the total number of photons $\bar{N}$. Obviously, from
Fig. 5 we can see that almost the same phase uncertainties can be obtained
in the limit of $\varphi \rightarrow 0$. In addition, when the values of
both $\bar{n}_{a}$ and $\bar{n}_{s}$ increase, these phase uncertainties
more quickly approach to the HL. Of course, when $\bar{n}_{a}=\bar{n}_{z}$
(or $\bar{n}_{s}=\bar{n}_{z}$), the optimal phase uncertainties will be
obtained. In Fig. 6, we repeat these graphs for $\varphi =0.015$. In the
latter case especially, we see that the optimal state is the PASVS. When $%
\bar{n}_{a}$\ and $\bar{n}_{s}$\ increase, the phase uncertainties blow up
due to the periodic nature of the expectation value of the parity operator,
but there are other photon numbers where the uncertainty is still below the
SNL for both the PASVS and the PSSVS.
\begin{figure}[tbph]
\centering \includegraphics[width=8cm]{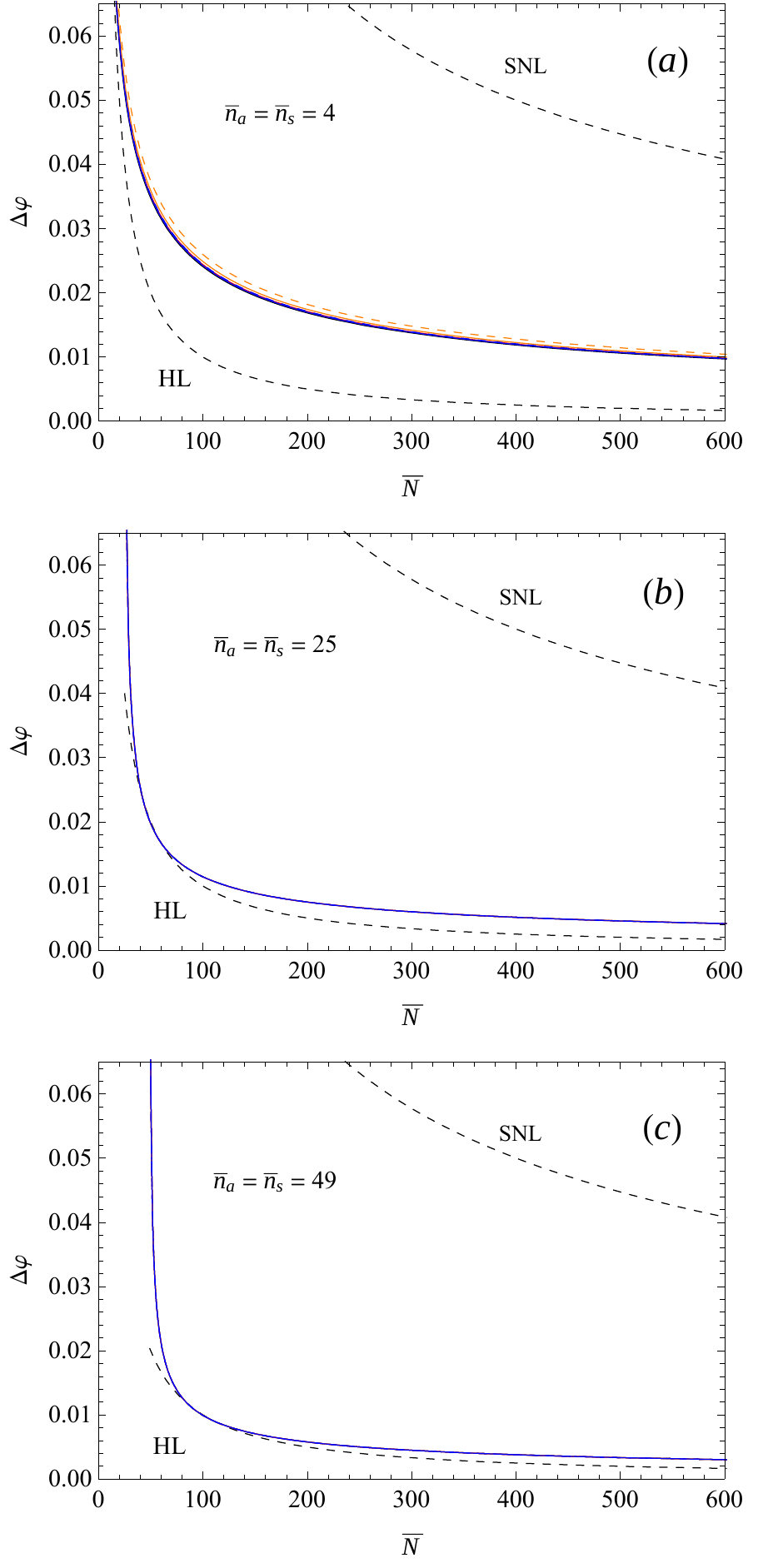}
\caption{Plots of the phase uncertainty versus total average photon at $%
\protect \varphi =10^{-4}$ for fixed the average photon number of the PASVS
and the PSSVS, as well as the SVS, along with the corresponding curves for
the SNL and the HL limits. Only the parameter $¦Á $ is being changed.}
\end{figure}
\begin{figure}[tbph]
\centering \includegraphics[width=8cm]{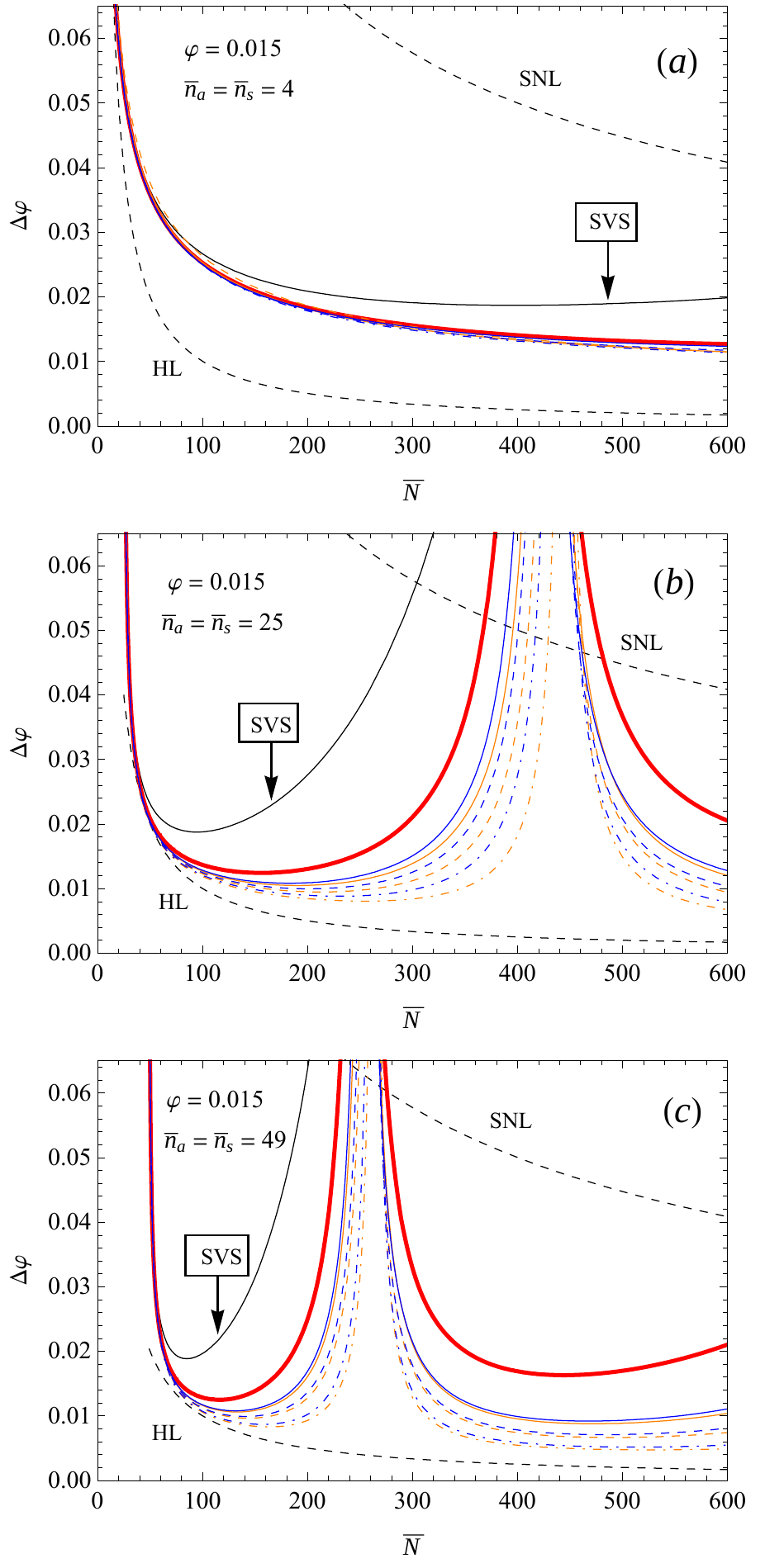}
\caption{Plots of the phase uncertainty versus the total average photon at $%
\protect \varphi =0.015$ for fixed the average photon number of the PASVS and
the PSSVS, as well as the SVS, along with the corresponding curves for the
SNL and the HL limits. The thick red line represents the PASVS(or
PSSVS)-coherent state with $k=l=1$; the orange line represents the
PASVS-coherent state with $k=2$; the orange dashed line denotes the
PASVS-coherent state with $k=3$; the orange dotted-dashed line denotes $k=6$%
. While, the blue line represents the PSSVS-coherent state with $l=2$; the
blue dashed line represents $l=3$; the blue dotted-dashed line denotes $l=6$%
. }
\end{figure}

Finally, we further investigate how the photon addition and subtraction
affect the phase sensitivity for a given initial squeezing parameter of the
SVS. Comparing with that results in Ref. \cite{21}, we find that, for the
same initial squeezing parameter $r$, it is also better to perform photon
addition rather than photon subtraction in respect to reducing the phase
uncertainty. This is because that it is always better to perform photon
addition rather than photon subtraction in order to increase the average
photon number for given the initial squeezing as shown in Fig. 1.

\begin{figure}[tbph]
\centering \includegraphics[width=8cm]{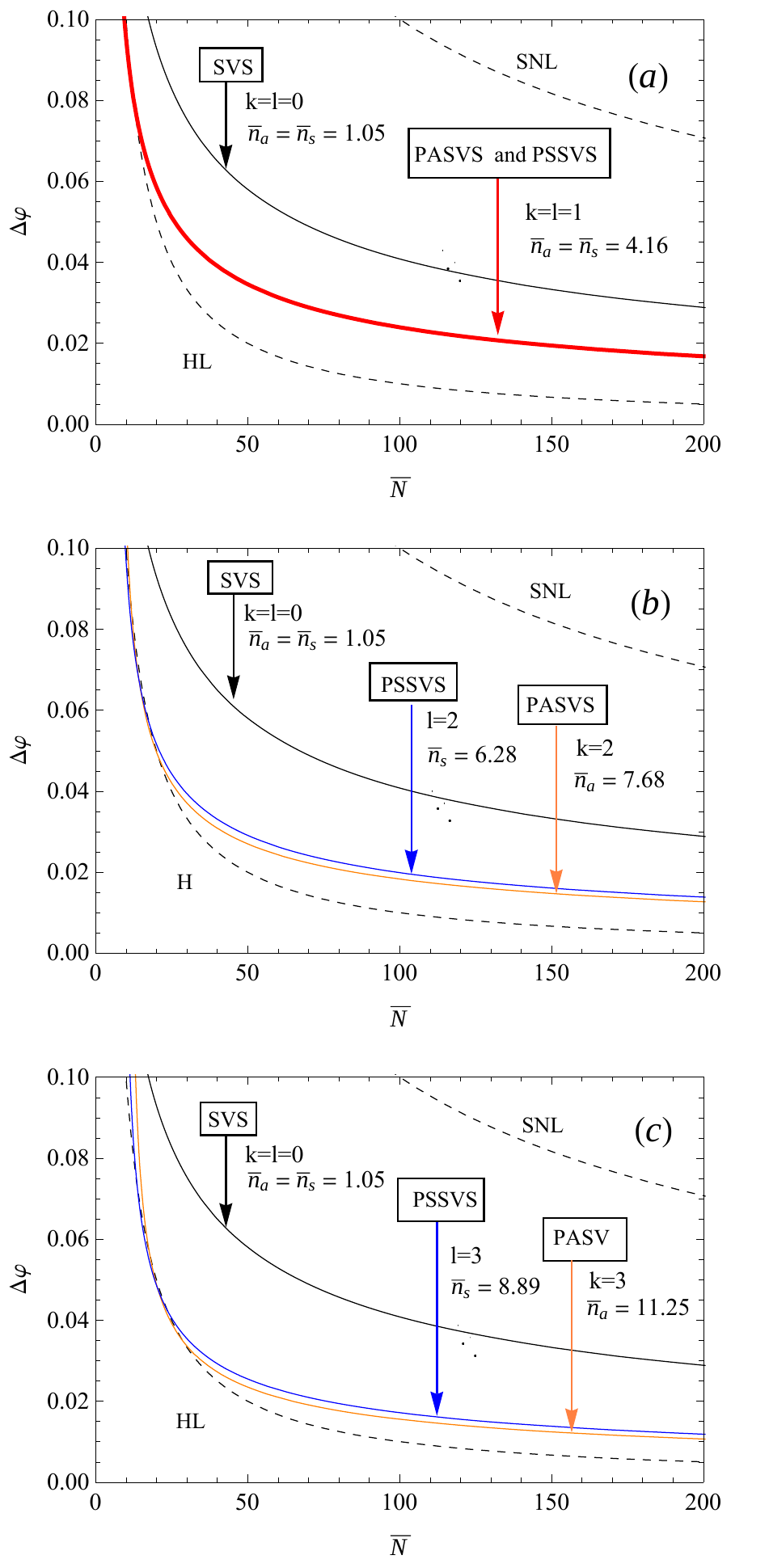}
\caption{For $r=0.9$, plots of the phase uncertainty versus the total
average photon number at $\protect \varphi =10^{-4}$, along with the
corresponding curves for the SNL and HL limits. Only the parameter $¦Á $
is being changed.}
\end{figure}

\section{Quantum fisher information of the MZI interferometer}

In this section, we will prove that the parity detection is the optimal
measurement for our considered interferometric scheme. Now, we use the
quantum Fisher information to find the maximum level of the phase
sensitivity by the Cram\'{e}r--Rao bound as given by \cite{17}%
\begin{equation}
\Delta \varphi _{\min }=\frac{1}{\sqrt{F_{Q}}},  \label{35}
\end{equation}%
For pure states injected into a MZI, the quantum Fisher information $F_{Q}$
can be obtained by \cite{r11}
\begin{equation}
F_{Q}=4\left[ \left \langle \psi ^{\prime }\left( \varphi \right) \right \vert
\left. \psi ^{\prime }\left( \varphi \right) \right \rangle -\left \vert
\left \langle \psi ^{\prime }\left( \varphi \right) \right \vert \left. \psi
\left( \varphi \right) \right \rangle \right \vert ^{2}\right] ,  \label{36}
\end{equation}%
where $\left \vert \psi \left( \varphi \right) \right \rangle =e^{-i\varphi
J_{3}}e^{-i\pi J_{1}/2}\left \vert \psi \right \rangle _{\text{in}}$ is the
state just before the second beam splitter of the MZI, and $\left \vert \psi
^{\prime }\left( \varphi \right) \right \rangle =\partial \left \vert \psi
\left( \varphi \right) \right \rangle /\partial \varphi $. In terms of the
input state, the quantum Fisher information becomes%
\begin{equation}
F_{Q}=4\left[ \left \langle \text{in}\right \vert J_{2}^{2}\left \vert \text{in}%
\right \rangle -\left \vert \left \langle \text{in}\right \vert J_{2}\left \vert
\text{in}\right \rangle \right \vert ^{2}\right] ,  \label{37}
\end{equation}%
and thus the quantum Fisher information is, up to factor of 4, the variance
of the operator $J_{2}$.

In the present work, the PASVS-coherent state as the interferometer state.
In order to obtain the quantum Fisher information, based on Eqs. (\ref{2})
and (\ref{3}), we first obtain the following expectation values, i.e., $%
\left \langle b\right \rangle _{\text{PASVS}}=\left \langle b^{\dagger
}\right \rangle _{\text{PASVS}}=0$, and
\begin{eqnarray}
N_{k+2,k} &\equiv &\left \langle b^{2}\right \rangle _{\text{PASVS}}  \notag
\\
&=&\frac{\partial ^{2k+2}}{\partial t^{k+2}\partial \tau ^{k}}e^{-\frac{%
\sinh 2r}{4}\left( t^{2}+\tau ^{2}\right) +t\tau \cosh ^{2}r}|_{t,\tau =0,}
\label{38}
\end{eqnarray}%
as well as%
\begin{eqnarray}
N_{k,k+2} &\equiv &\left \langle b^{\dagger 2}\right \rangle _{\text{PASVS}}
\notag \\
&=&\frac{\partial ^{2k+2}}{\partial t^{k}\partial \tau ^{k+2}}e^{-\frac{%
\sinh 2r}{4}\left( t^{2}+\tau ^{2}\right) +t\tau \cosh ^{2}r}|_{t,\tau =0}.
\label{39}
\end{eqnarray}%
Then, according to Eqs. (\ref{10}) and (\ref{37}), we can directly obtain
the Quantum Fisher information of the MZI as%
\begin{equation}
F_{QA}=2\bar{n}_{z}\bar{n}_{a}+\bar{n}_{z}+\bar{n}_{a}-2\bar{n}_{z}\frac{%
N_{k+2,k}}{N_{k}},  \label{40}
\end{equation}%
where we have also set $\theta =0$ (the phase of the coherent state) for
obtaining the good phase uncertainty and used the relation $%
N_{k+2,k}=N_{k,k+2}$. For $k=0,1,2$, combining Eqs. (\ref{31}-\ref{34}) and
Eq. (\ref{40}), we can analytically prove that the quantum Cram\'{e}r-Rao
bound can be reached via the parity detection in the limit $\varphi
\rightarrow 0$. For general $k$, we can numerically prove this is still true.

Using the similar method, we derive the quantum Fisher information for the
PSSVS-coherent interferometer state

\begin{equation}
F_{QS}=2\bar{n}_{z}\bar{n}_{s}+\bar{n}_{z}+\bar{n}_{s}-2\bar{n}_{z}\frac{%
C_{l+2,l}}{C_{l}},  \label{41}
\end{equation}%
where
\begin{eqnarray}
C_{l+2,l} &\equiv &\left \langle b^{2}\right \rangle _{\text{PSSVS}}  \notag \\
&=&\frac{\partial ^{2k+2}}{\partial t^{k+2}\partial \tau g^{k}}e^{-\frac{%
\sinh 2r}{4}\left( t^{2}+\tau ^{2}\right) +t\tau \sinh ^{2}r}|_{t,\tau =0,}
\notag \\
C_{l,l+2} &\equiv &\left \langle b^{\dagger 2}\right \rangle _{\text{PSSVS}}
\notag \\
&=&\frac{\partial ^{2k+2}}{\partial t^{k}\partial \tau ^{k+2}}e^{-\frac{%
\sinh 2r}{4}\left( t^{2}+\tau ^{2}\right) +t\tau \sinh ^{2}r}|_{t,\tau =0},
\label{42}
\end{eqnarray}%
and we have used the relation $C_{l+2,l}=C_{l,l+2}$. Similarly, we can check
that the quantum Cram\'{e}r-Rao bound can be reached via the parity
detection in the limit $\varphi \rightarrow 0$ for the PSSVS-coherent
considered as the interferometer state.

\section{Conclusions}

In summary, we have studied the quantum optimal interference by mixing a
coherent state with a PASVS. Given a constraint on the total average number
of photons and in the limit of $\varphi \rightarrow 0$, for a MZI with
PASVS-coherent and PSSVS-coherent as well as SVS-coherent input states,
almost the same phase uncertainties can be obtained. However, when the phase
shift $\varphi $\ somewhat deviates from zero, the optimal state is neither
the SVS nor the PSSVS, but the PASVS when these three states contain many
photons. On the other hand, for fixed the initial squeezing $r$, it is
better to perform photon addition rather than photon subtraction for
improving the phase sensitivity of the MZI. This may be because that it is
always better to perform addition rather than subtraction in order to
increase the average photon number of the SVS for given the initial
squeezing. Finally, we show that the quantum Cram\'{e}r-Rao bound can be
reached via the parity detection in the limit $\varphi \rightarrow 0$.

\section*{Acknowledgments}

This work is supported by the National Natural Science Foundation of China
(No. 11690032, No. 11665013, and No. 11704051) and sponsored by Qing Lan
Project of the Higher Educations of Jiangsu Province of China.

\end{document}